%
%
%
%
%
%
%
%
%
%
\hoffset=0.0in
\voffset=0.0in
\hsize=6.5in
\vsize=8.9in
\normalbaselineskip=12pt
\normalbaselines
\topskip=\baselineskip
\parindent=15pt
%
%
%

\let\dl=\delta

\let\la=\langle
\let\ra=\rangle

\let\lf=\left
\let\rt=\right
\let\dt=\cdot
\let\del=\nabla
\let\dg=\dagger

\let\q=\widehat
\let\Q=\overbrace
\let\h=\hbar

\let\rta=\rightarrow

\let\x=\times
\let\dy=\displaystyle

\let\sy=\scriptstyle

\let\:=\>
\let\\=\cr
\let\emph=\e

\let\m=\hbox

\let\cl=\centerline

\def\e#1{{\it #1\/}}
\def\textbf#1{{\bf #1}}
\def\[{$$}
\def\]{\[}
\def\re#1#2{$$\matrix{#1\cr}\eqno{({\rm #2})}$$}
\def\de#1{$$\matrix{#1\cr}$$}

\def\eqdf{\buildrel{\rm def}\over =}
\def\hf{{\sy {1 \over 2}}}

\def\qH{\q{H}}

\def\mathrm#1{{\rm #1}}
\def\mr#1{{\rm #1}}
\def\mathcal#1{{\cal #1}}

\def\Hclc{H_\mr{cl}([\vx, \vp], t)}
\def\Hclcp{H_\mr{cl}([\vx', \vp'], t)}
\def\qHcnc{\qH([\qvx, \qvp], t)}
\def\cs{\mr{cl}}
\def\Hclcs{H_\mr{cl}([\vx_\cs(t),\, \vp_\cs(t)], t)}

\def\mbf{\fam\bffam\tenbf}
\def\bv#1{{\mbf #1}}

\def\vx{\bv{x}}
\def\vy{\bv{y}}

\def\vv{\bv{v}}

\def\vp{\bv{p}}

\def\vj{\bv{j}}
\def\vk{\bv{k}}
\def\vl{\bv{l}}
\def\vz{\bv{0}}

\def\qvx{\q{\vx}}
\def\qvp{\q{\vp}}

\def\qp{\q{p}}
\def\qx{\q{x}}

\def\vjc{\vj(\vx, t;[|\psi(t)\ra, \qHcnc])}
\def\vjcy{\vj(\vy, t;[|\psi(t)\ra, \qHcnc])}
\def\dvjc{\del_\vx\dt\vjc}
\def\dvjcy{\del_\vy\dt\vjcy}

\def\Schr{Schr\"{o}\-ding\-er}
\font\frtbf = cmbx12 scaled \magstep1
\font\twlbf = cmbx12
\font\ninbf = cmbx9
\font\svtrm = cmr17
\font\twlrm = cmr12
\font\ninrm = cmr9
\font\ghtrm = cmr8

\def\gr#1{{\ghtrm #1}}

\def\abstract#1{{\ninbf\cl{Abstract}}\medskip
\openup -0.1\baselineskip
{\ninrm\leftskip=2pc\rightskip=2pc\noindent #1\par}
\normalbaselines}
\def\sct#1{\vskip 1.33\baselineskip\noindent{\twlbf #1}\medskip}

\def\so{\raise 0.65ex \m{\sevenrm 1}}
\def\sk{\par\vskip 0.66\baselineskip}
{\svtrm
\cl{Obtaining the Probability Vector Current Density in}
\medskip
\cl{Canonical Quantum Mechanics by Linear Superposition}
}
\bigskip
{\twlrm
\cl{Steven Kenneth Kauffmann\footnote{${}^\ast$}{\gr{Retired, American
Physical Society Senior Life Member, E-mail: SKKauffmann@gmail.com}}}
}
\bigskip\smallskip
\abstract{%
The quantum mechanics status of the probability vector current density
has long seemed to be marginal.  On one hand no systematic prescrip%
tion for its construction is provided, and the special examples of it
that are obtained for particular types of Hamiltonian operator could
conceivably be attributed to happenstance.  On the other hand this
concept's key physical interpretation as local average particle flux,
which flows from the equation of continuity that it is supposed to
satisfy in conjunction with the probability scalar density, has been
claimed to breach the uncertainty principle.  Given the dispiriting
impact of that claim, we straightaway point out that the subtle direc%
tional nature of the uncertainty principle makes it consistent with
the measurement of local average particle flux.  We next focus on the
fact that the unique closed-form linear-superposition quantization
of any classical Hamiltonian function yields in tandem the correspond%
ing unique linear-superposition closed-form divergence of the proba%
bility vector current density.  Because the probability vector current
density is linked to the quantum physics only through the occurrence
of its divergence in the equation of continuity, it is theoretically
most appropriate to construct this vector field exclusively from its
divergence---analysis of the best-known ``textbook'' special example
of a probability vector current density shows that it is thus con%
structed.  That special example in fact leads to the physically inter%
esting ``Ehrenfest subclass'' of probability vector current densities,
which are closely related to their classical peers.
}

\sct{Introduction}
\noindent
The quantum mechanical probability vector current density concept has
long been at best hazily understood.  Although special examples of
probability vector current density have been obtained for particular
types of Hamiltonian operator~[1], there is no systematic prescription
for constructing it, so its special examples could conceivably be at%
tributed to happenstance.  Its essential feature is supposed to be
that, in consequence of the quantum mechanical conservation of proba%
bility, it jointly with the probability scalar density satisfies the
equation of continuity, which compellingly suggests that it physically
represents local average particle flux.  It has, however, been claimed
that the measurement of local average particle flux breaches the un%
certainty principle because such a measurement involves arbitrary par%
ticle localization while it simultaneously yields information concern%
ing particle velocity, and for that reason probability vector current
density \e{cannot have its only truly natural physical interpreta%
tion}, namely that of local average particle flux~[1].  This conten%
tion, if valid, would raise two linked vexing issues: (1) if probabil%
ity vector current density cannot in principle be physically inter%
preted as local average particle flux, then \e{what} is its correct
physical interpretation, and (2) is that correct physical interpreta%
tion still something that is of actual interest to physicists?

Very forturnately, however, while it is indeed the case that measure%
ment of local average particle flux restricts particle position while
delivering information regarding particle velocity, these things do
not in fact occur in such a manner as to challenge the uncertainty
principle.  To measure one of the vector components of local average
particle flux at a given point, one passes a plane which is perpendic%
ular to that vector component through that point.  One then selects an
arbitrarily small region of that plane centered on that point and
measures the average rate that particles pass through that selected
planar region.  This average rate, divided by the area of the selected
planar region, is an approximation to that particular vector component
of average particle flux at that particular point, an approximation
which is, in  principle, refined by additionally shrinking the planar
region on which that point is centered.  The flux one thus measures
does indeed reflect the component of average particle velocity in the
direction \e{perpendicular} to the plane, provided, of course, that
one knows the average particle density. The uncertainty principle,
however, \e{does not restrict the accuracy} with which a particular
component of particle momentum can be determined due to restrictions
that are imposed on the particle's position in directions which are
\e{perpendicular} to that momentum component.  To see this one need
look no further than the Dirac canonical commutation rule,
\de{[\qx_i, \qx_j] = 0,\enskip [\qp_i, \qp_j] = 0,\enskip
    [\qx_i, \qp_j] = i\h\dl_{ij},\enskip i,j = 1, 2, 3,}
and recall that the uncertainty principle \e{only applies} to quan%
tized dynamical variable pairs \e{that fail to commute}, which is
clearly \e{not the case} for a position component and a momentum com%
ponent \e{that are mutually perpendicular}.  Therefore \e{the measure%
ment of local average particle flux}, notwithstanding that it res%
tricts particle position while delivering averaged particle velocity
information---namely in the direction \e{perpendicular} to the plane
in which particle position is restricted---\e{does not challenge the
uncertainty principle}.

Thus there is in fact no known valid reason to refrain from physically
interpreting probability vector current density as local average par%
ticle flux, even as the equation of continuity ostensibly satisfied by
probability vector current density in conjunction with probability
scalar density so cogently suggests is sound.  The bona fide issue
here is a \e{different} one, namely the lack of a \e{general} pres%
cription for \e{constructing} that desired probability vector current
density which \e{actually satisfies} the equation of continuity in
conjunction with the probability scalar density.

We therefore in the rest of this article concentrate on systematically
working out such a probability vector current density for any canoni%
cal Hamiltonian operator whatsoever (at least in principle---as one
can well imagine, anything worked out to such a degree of generality
yields formidable expressions that cannot be expected to have trans%
parency as their strong suit).  We begin with the \e{divergence} of
the probability vector current density, which, from the equation of
continuity, must be equal to the negative of the time derivative of
\e{the well-defined probability scalar density.}  That time deriva%
tive, in turn, is the state-vector expectation value of $(i/\h)$ times
the commutator of the set of position projection operators \e{with the
Hamiltonian operator}.  Because there exists a linear superposition
technique for the unique closed-form quantization of \e{any classical
Hamiltonian function}~[2], it turns out that for Hamiltonian operators
which have classical Hamiltonian function antecedents we can \e{also}
uniquely reduce the just-mentioned expression for the \e{divergence}
of the probability vector current density, which involves the state-%
vector expectation value of a \e{Hamiltonian-operator} commutator, to
a linear-superposition closed form.  Thus the \e{divergence} of the
probability vector current density can always be fully and uniquely
worked out when the Hamiltonian operator has a classical Hamiltonian
function antecedent, i.e., when it is a \e{canonical} Hamiltonian op%
erator.

The  probability vector current density \e{itself} of course \e{is
naturally mathematically ambiguous}, e.g., it tolerates \e{the addi%
tion of an arbitrary vector field that is a curl}.  However, since its
\e{divergence} is a state-vector \e{expectation value} (of a set of
operators), it is physically entirely reasonable to \e{restrict} the
probability vector current density to being a \e{homogeneously linear
functional} of that \e{divergence} so that \e{it likewise will be a
state-vector expectation value}.  Indeed, the larger idea of \e{res%
tricting} the probability vector current density to the very barest
\e{minimum} that is compatible with its divergence makes impeccable
sense because it is \e{only that divergence} which makes \e{actual
contact with the system's quantum physics}, doing so, of course,
through the \e{equation of continuity} that it satisfies in conjunc%
tion with the probability scalar density.  Therefore the probability
vector current density is devoid of quantum physical information
\e{that is not already implicit in its divergence}.  We shall formally
\e{implement} this crucial point by stipulating not only that the
probability current density is \e{homogeneously linear} in its diver%
gence, but that it furthermore \e{must not} depend on any \e{con%
stants} which are \e{additional} to those that are \e{intrinsic to its
divergence}, and also that among the forms which are mathematical can%
didates for this \e{vector} field, \e{only the most symmetric are to
be considered} because these \e{add no further information} to that
available from its \e{scalar} divergence. These three stipulations re%
sult in a \e{unique closed-form expression} for the probability vector
current density in terms of its unique divergence, and we shall see
that the ``textbook'' best-known special example of a probability vec%
tor current density~[1] is indeed consistent with these three stipula%
tions.

We now turn to the presentation of \e{the ingredients that are needed}
to work out the probability vector current density in the manner that
has just been outlined.  As is well-known, the \e{conservation of
probability} in quantum mechanics~[1] follows from the Hermitian prop%
erty of the Hamiltonian operator, namely $\qH^\dg(t) = \qH(t)$, in
conjunction with the \Schr\ equation,
\re{
    i\h d|\psi(t)\ra/dt = \qH(t)|\psi(t)\ra,
}{1a}
which together imply that,
\re{
    -i\h d\la\psi(t)|/dt = \la\psi(t)|\qH(t),
}{1b}
and therefore that probability is conserved,
\re{
    d\la\psi(t)|\psi(t)\ra/dt = \la\psi(t)|\lf[(i/\h)\qH(t) +
                                (-i/\h)\qH(t)\rt]|\psi(t)\ra = 0.
}{1c}

In the case of \e{canonical} quantum mechanics, $\qH(t)$ is \e{unique%
ly obtained} from a \e{classical Hamiltonian function} $\Hclc$ by the
imposition of the following \e{self-consistent} slight extension of
Dirac's canonical commutation rule~[2],
\re{
    [f_1(\qvx) + g_1(\qvp), f_2(\qvx) + g_2(\qvp)] =
    i\h\,\Q{\,\{f_1(\vx) + g_1(\vp),\, f_2(\vx) + g_2(\vp)\}\,}\, ,
}{2}
where the $\{\enskip,\enskip\}$ are the \e{classical Poisson brack%
ets}, and the overbrace is used to denote \e{quantization}---the over%
brace is \e{only} used here for that purpose because the traditional
hat accent for denoting quantization \e{is not sufficiently exten%
sible}.  The Eq.~(2) slight extension of Dirac's canonical commutation
rule turns out to self-consistently \e{completely resolve} the Dirac
\e{operator-ordering ambiguity} in favor of Born-Jordan operator or%
dering~[2], which is \e{exactly the same operator ordering that is im%
plicit in the Hamiltonian path integral}~[3].  The \e{unique} Hamil%
tonian operator which follows from the classical Hamiltonian function
$\Hclc$ via the application of Eq.~(2) is, of course, denoted as $\Q{
\,\Hclc\,}\,$, or, when \e{explicit reference} to the underlying clas%
sical Hamiltonian function $\Hclc$ \e{isn't necessary}, as simply the
\e{canonical Hamiltonian operator} $\qHcnc$.

When such a \e{canonical} $\qHcnc$ describes the quantum mechanics,
probability conservation, given by Eq.~(1c), can be expressed in terms
of the \e{probability scalar density} $(\la\psi(t)|\vx\ra\la\vx|\psi
(t)\ra)$ because of the following expansion in the complete set of
position states $|\vx\ra$,
\re{
    d\la\psi(t)|\psi(t)\ra/dt =
    (d/dt)\int\la\psi(t)|\vx\ra\la\vx|\psi(t)\ra\,d^N\vx = 0,
}{3a}
which relation might plausibly be expected to imply that the time
derivative $d(\la\psi(t)|\vx\ra\la\vx|\psi(t)\ra)/dt$ of the probabil%
ity scalar density $(\la\psi(t)|\vx\ra\la\vx|\psi(t)\ra)$ \e{is a per%
fect differential in the vector variable} $\vx$, namely that there ex%
ists \e{a probability vector current density} $\vjc$ which satisfies,
\re{
    d(\la\psi(t)|\vx\ra\la\vx|\psi(t)\ra)/dt + \dvjc = 0.
}{3b}
Eq.~(3b) is the \e{equation of continuity} that the probability vector
current density $\vjc$ \e{is supposed to satisfy} in conjunction with
the probability scalar density $(\la\psi(t)|\vx\ra\la\vx|\psi(t)\ra)$
as a plausible consequence of the conservation of probability that is
given by Eq.~(3a).  In light of the \Schr\ equations given by Eqs.~%
(1a) and (1b), as they apply to the particular case that $\qH(t) =
\qHcnc$, we can see that \e{actually fulfilling the} Eq.~(3b) \e{equa%
tion of continuity requires that},
\re{
    \dvjc = - d(\la\psi(t)|\vx\ra\la\vx|\psi(t)\ra)/dt =\cr
    (i/\h)\lf[\la\psi(t)|\vx\ra\la\vx|\qHcnc|\psi(t)\ra -
           \la\psi(t)|\qHcnc|\vx\ra\la\vx|\psi(t)\ra\rt] =\cr
    \Re\!\lf[(2i/\h)\la\psi(t)|\vx\ra\la\vx|\qHcnc|\psi(t)\ra\rt],
}{3c}
where the symbol $\Re$ that occurs after the last equal sign of Eq.~%
(3c) denotes the real part of the bracketed expression which follows
it.  We thus see that to obtain the \e{divergence} of $\vjc$ we need
to evaluate the \e{core part} $\la\vx|\qHcnc|\psi(t)\ra$ of that
bracketed expression for an \e{arbitrary} canonical Hamiltonian opera%
tor $\qHcnc$.  To \e{achieve} this goal it will be extremely useful to
emulate the procedure of Ref.~[2] wherein such an \e{arbitrary}
$\qHcnc$ is \e{linearly decomposed} into ``Fourier component'' \e{op%
erators} of the form $e^{\mp i\qvx\dt\vk}e^{\pm i\qvp\dt\vl/\h}$ be%
cause, of course,
\re{
    \la\vx|e^{\mp i\qvx\dt\vk}e^{\pm i\qvp\dt\vl/\h}|\psi(t)\ra =
    e^{\mp i\vx\dt\vk}\la\vx|e^{\pm i\qvp\dt\vl/\h}|\psi(t)\ra =
    e^{\mp i\vx\dt\vk}\la\vx\pm\vl|\psi(t)\ra.
}{3d}
\vfil
\break

\sct{The probability vector current density divergence by linear
superposition}
\noindent
Following Ref.~[2], we note that the key step for decomposing $\qHcnc$
into ``Fourier component'' \e{operators} is the \e{orthodox} corres%
ponding Fourier decomposition of its \e{underlying classical Hamilton%
tonian function} $\Hclc$.  This, of course, follows from the identity,
\re{
    \Hclc = \int d^N\vx'\,\int d^N\vp'\>\dl^{(N)}\!\!(\vx - \vx')\;
    \dl^{(N)}\!\!(\vp - \vp')\>\Hclcp,
}{4a}
after we insert into it the Fourier delta-function representations,
\re{
    \dl^{(N)}\!\!(\vx - \vx') =
    (2\pi)^{-N}\int d^N\vk\; e^{-i\vk\dt(\vx - \vx')},\enskip
    \dl^{(N)}\!\!(\vp - \vp') =
    (2\pi\h)^{-N}\int d^N\vl\; e^{i\vl\dt(\vp - \vp')/\h}.
}{4b}
Since $\Hclc$ is a real-valued function, after inserting Eq.~(4b) into
Eq.~(4a) it is convenient to furthermore explicitly discard the van%
ishing imaginary part of the result, which yields,
\re{
    \Hclc = (4\pi^2\h)^{-N}\int d^N\vx'\,\int d^N\vp'\,\Hclcp\,
    \int d^N\vk\,\int d^N\vl\;\x\cr
    \bigl(\cos(-\vx'\dt\vk + \vp'\dt\vl/\h)\cos(-\vx\dt\vk + \vp\dt\vl/\h) +
     \sin(-\vx'\dt\vk + \vp'\dt\vl/\h)\sin(-\vx\dt\vk + \vp\dt\vl/\h)\bigr).
}{4c}
The fact that quantization is a \e{linear process}~[2] permits us to
conclude from Eq.~(4c) that,
\re{
    \qHcnc = \Q{\,\Hclc\,} = (4\pi^2\h)^{-N}
    \int d^N\vx'\,\int d^N\vp'\,\Hclcp\,\int d^N\vk\,\int d^N\vl\;\x\cr
    \Bigl(\cos(-\vx'\dt\vk + \vp'\dt\vl/\h)\,
    \Q{\,\cos(-\vx\dt\vk + \vp\dt\vl/\h)\,}\,+\,
    \sin(-\vx'\dt\vk + \vp'\dt\vl/\h)\,
    \Q{\,\sin(-\vx\dt\vk + \vp\dt\vl/\h)\,}\,
    \Bigr).
}{4d}
As in Ref.~[2], we obtain $\,\Q{\,\exp(-i\vx\dt\vk + i\vp\dt\vl/\h)\,}
\,$ from Eq.~(2).  From this we further immediately obtain $\,\Q{\,
\exp(i\vx\dt\vk - i\vp\dt\vl/\h)\,}\,$, and those two results together
yield $\,\Q{\,\cos(-\vx\dt\vk + \vp\dt\vl/\h)\,}\,$ and $\,\Q{\,\sin(
-\vx\dt\vk + \vp\dt\vl/\h)\,}\,$, which is what we require for inser%
tion into Eq.~(4d).  We shall, however, \e{insure} that these cosine
and sine quantizations are expressed as linear combinations of prod%
ucts of exponential operators of the ``Fourier component'' type that
were introduced in Eq.~(3d), because it is \e{those} ``Fourier compo%
nent'' type operators which are \e{transparently useful} for the eval%
uation of the \e{core part} of the probability vector current densi%
ty's \e{divergence} $\dvjc$, as is apparent from Eqs.~(3c) and (3d).

We now carry out the Eq.~(2) quantization of $\exp(-i\vx\dt\vk + i\vp
\dt\vl/\h)$ with the goal of expressing its result in terms of Eq.~%
(3d) ``Fourier component'' type operators.  With the needed exponen%
tial ingredients inserted, Eq.~(2) reads,
\re{
    [e^{-i\qvx\dt\vk}, e^{i\qvp\dt\vl/\h}] = i\h\,\Q{\,
    \{e^{-i\vx\dt\vk},\, e^{i\vp\dt\vl/\h}\}\,}\, ,
}{4e}
which, written out, is,
\re{
    e^{-i\qvx\dt\vk} e^{i\qvp\dt\vl/\h} - e^{i\qvp\dt\vl/\h}e^{-i\qvx\dt\vk} =
    i(\vk\dt\vl)\,\Q{\,e^{-i\vx\dt\vk + i\vp\dt\vl/\h}\,}\, .
}{4f}
Eq.~(4f) can then be reexpressed as,
\re{
    \lf(e^{i(\vk\dt\vl)/2} - e^{-i(\vk\dt\vl)/2}\rt)
    e^{-i\qvx\dt\vk + i\qvp\dt\vl/\h}
    = i(\vk\dt\vl)\,\Q{\,e^{-i\vx\dt\vk + i\vp\dt\vl/\h}\,}\, .
}{4g}
which yields the unique exponential quantization,
\re{
    \Q{\,e^{-i\vx\dt\vk + i\vp\dt\vl/\h}\,} =
    e^{-i\qvx\dt\vk + i\qvp\dt\vl/\h}\sin(\hf\vk\dt\vl)/(\hf\vk\dt\vl) =
    e^{-i\qvx\dt\vk}e^{i\qvp\dt\vl/\h}e^{-i\hf\vk\dt\vl}\sin(\hf\vk\dt\vl)/
    (\hf\vk\dt\vl),
}{4h}
where the form farthest to the right in Eq.~(4h) is expressed in terms
of the desired ``Fourier component'' type operator.  We now make the
simple substitutions $\vk\rta -\vk$ and $\vl\rta -\vl$ in Eq.~(4h),
which produce the additional useful result that,
\re{
    \Q{\,e^{i\vx\dt\vk - i\vp\dt\vl/\h}\,} =
    e^{i\qvx\dt\vk - i\qvp\dt\vl/\h}\sin(\hf\vk\dt\vl)/(\hf\vk\dt\vl) =
    e^{i\qvx\dt\vk}e^{-i\qvp\dt\vl/\h}e^{-i\hf\vk\dt\vl}\sin(\hf\vk\dt\vl)/
    (\hf\vk\dt\vl).
}{4i}
Combining Eqs.~(4h) and (4i) then yields,
\re{
    \Q{\,\cos(-\vx\dt\vk + \vp\dt\vl/\h)\,} =
    (1/2)\lf(e^{-i\qvx\dt\vk}e^{i\qvp\dt\vl/\h} +
    e^{i\qvx\dt\vk}e^{-i\qvp\dt\vl/\h}\rt)
    e^{-i\hf\vk\dt\vl}\sin(\hf\vk\dt\vl)/(\hf\vk\dt\vl),
}{4j}
and,
\re{
    \Q{\,\sin(-\vx\dt\vk + \vp\dt\vl/\h)\,} =
    (1/(2i))\lf(e^{-i\qvx\dt\vk}e^{i\qvp\dt\vl/\h} -
    e^{i\qvx\dt\vk}e^{-i\qvp\dt\vl/\h}\rt)
    e^{-i\hf\vk\dt\vl}\sin(\hf\vk\dt\vl)/(\hf\vk\dt\vl),
}{4k}
which are expressed in terms of the desired ``Fourier component'' type
operators.  We can now insert Eqs.~(4j) and (4k) into Eq.~(4d), which
in turn is inserted into Eq.~(3c), after which Eq.~(3d) is straight%
forwardly applied.  The upshot is the general \e{linear superposition
expression} for the \e{divergence} of the probability vector current
density that results from an \e{arbitrary canonical Hamiltonian opera%
tor} $\qHcnc$ \e{whose classical antecedent Hamiltonian function is}
$\Hclc$,
\re{
    \dvjc =\cr (4\pi^2\h)^{-N}\int d^N\vx'\,\int d^N\vp'\,
    (\Hclcp/\h)\,\int d^N\vk\,\int d^N\vl\>
    \lf(\sin(\hf\vk\dt\vl)/(\hf\vk\dt\vl)\rt)\,\x\cr
    \Bigl(\cos(-\vx'\dt\vk + \vp'\dt\vl/\h)\,
\Re\!\bigl[ie^{-i\hf\vk\dt\vl}\la\psi(t)|\vx\ra
\lf(e^{-i\vx\dt\vk}\la\vx+\vl|\psi(t)\ra +
e^{i\vx\dt\vk}\la\vx-\vl|\psi(t)\ra\rt)\bigr] +\cr
     \sin(-\vx'\dt\vk + \vp'\dt\vl/\h)\,
\Re\!\bigl[e^{-i\hf\vk\dt\vl}\la\psi(t)|\vx\ra
\lf(e^{-i\vx\dt\vk}\la\vx+\vl|\psi(t)\ra -
e^{i\vx\dt\vk}\la\vx-\vl|\psi(t)\ra\rt)\bigr]\Bigr).
}{5a}
One interesting \e{special case} of Eq.~(5a) occurs when $\Hclc$ has
no dependence on the configuration variables $\vx$.  In that case the
integrations over the variables $\vx'$ and $\vk$ that occur in Eq.~%
(4d) are obviously superfluous, and, if actually carried out in Eq.~%
(5a), simply eliminate $\vx'$ along with a factor of $(2\pi)^{-N}$,
while setting $\vk$ to $\vz$.  Thus in the case that $\Hclc$ has no
dependence on $\vx$, Eq.~(5a) simplifies to,
\re{
    \del_\vx\dt\vj(\vx, t;[|\psi(t)\ra, \qH([\qvp], t)]) =
    (2\pi\h)^{-N}\int d^N\vp'\,
    (H_\mr{cl}([\vp'], t)/\h)\,\int d^N\vl\;\x\cr
    \Bigl(\cos(\vp'\dt\vl/\h)\,
\Re\!\lf[i\la\psi(t)|\vx\ra
\lf(\la\vx+\vl|\psi(t)\ra + \la\vx-\vl|\psi(t)\ra\rt)\rt] +\cr
     \sin(\vp'\dt\vl/\h)\,
\Re\!\lf[\la\psi(t)|\vx\ra
\lf(\la\vx+\vl|\psi(t)\ra - \la\vx-\vl|\psi(t)\ra\rt)\rt]\Bigr).
}{5b}
Similarly, if $\Hclc$ has no dependence on the momentum variables $\vp
$, then the integrations over $\vp'$ and $\vl$ in Eq.~(5a) eliminate
$\vp'$ along with a factor of $(2\pi\h)^{-N}$, while setting $\vl$ to
$\vz$.  Thus in the case that $\Hclc$ has no dependence on $\vp$, Eq.%
~(5a) yields,
\re{
    \del_\vx\dt\vj(\vx, t;[|\psi(t)\ra, \qH([\qvx], t)]) =
    (2\pi)^{-N}\int d^N\vx'\,
    (H_\mr{cl}([\vx'], t)/\h)\,\int d^N\vk\;\x\cr
    \Bigl(\cos(-\vx'\dt\vk)\,
\Re\!\lf[i|\la\psi(t)|\vx\ra|^2
\lf(e^{-i\vx\dt\vk} + e^{i\vx\dt\vk}\rt)\rt] +\cr
     \sin(-\vx'\dt\vk)\,
\Re\!\lf[|\la\psi(t)|\vx\ra|^2
\lf(e^{-i\vx\dt\vk} - e^{i\vx\dt\vk}\rt)\rt]\Bigr) = 0,
}{5c}
namely that $\del_\vx\dt\vj(\vx, t;[|\psi(t)\ra, \qH([\qvx], t)])$
\e{vanishes identically}.  This can \e{also} be verified \e{directly}
from Eq.~(3c) for such a canonical Hamiltonian operator $\qH([\qvx],
t)$, because $\la\vx|$ is one of its \e{eigenvectors}, with the cor%
responding \e{real} eigenvalue $H_\mr{cl}([\vx], t)$, and $\la\psi(t)|
\vx\ra\la\vx|\psi(t)\ra$ is \e{real-valued} as well.  We therefore see
that \e{any terms of} $\Hclc$ \e{which have no dependence on} $\vp$
\e{can simply be truncated from} $\Hclc$ \e{before} $\Hclc$ \e{is in%
serted} into the general linear superposition expression of Eq.~(5a)
for the divergence of the probability vector current density $\dvjc$.

Our \e{ultimate} goal, of course, is the probability vector current
density $\vjc$ \e{itself}, rather than its divergence.  We now turn
our focus to obtaining it, recalling that in the Introduction we
set out three stipulations to be imposed on it to make it formally
consistent with the fact that it is devoid of physical information
that is not already implicit in its divergence.

\sct{Obtaining the probability vector current density from its diver%
gence}
\noindent
A homogeneously linear form in terms of its divergence for the $n$th
component of the probability vector current density is given by,
\re{
    \bigl(\vjc\bigr)_n =\cr w_n{\dy\int_{x_n^{(0)}}^{x_n}}
    \del_\vx\dt\vj(x_1,\ldots,x'_n,\ldots, x_N, t;
    [|\psi(t)\ra, \qHcnc])dx'_n,\enskip n = 1, \ldots, N,
}{6a}
where the \e{weights} $w_n$ satisfy $w_1 + \cdots + w_N = 1$.  We of
course stipulated \e{not only} that the probability vector current
density is homogeneously linear in its divergence, but \e{also} that
it has \e{no dependence} on constants which are \e{additional} to
those that are intrinsic to its divergence.  The \e{dependence} of the
expression on the right-hand side of Eq.~(6a) on the set of constants
$\{x_1^{(0)}, \ldots, x_N^{(0)}\}$ that are its \e{lower limits of in%
tegration} is readily \e{removed} by setting \e{all} those lower lim%
its of integration to $-\infty$.  That the result of thus introducing
an infinite integration interval is well-defined (i.e., does not di%
verge) is directly tied in with the conservation of probability, as
one can see from Eqs.~(3a) and (3b) (with the latter repeated in Eq.~%
(3c)).  Finally, we have \e{also} stipulated that among the forms of
the mathematical candidates for the probability vector current densi%
ty, \e{only the most symmetric are to be considered}. That stipula%
tion, and to a certain extent also the injunction against additional
constants not intrinsic to its divergence, requires us to set \e{every
weight value} $w_n$ of Eq.~(6a) to $1/N$, $n = 1, \ldots, N$.  With
these specifications, Eq.~(6a) becomes,
\re{
    \bigl(\vjc\bigr)_n =\cr (1/N){\dy\int_{-\infty}^{x_n}}
    \del_\vx\dt\vj(x_1,\ldots,x'_n,\ldots, x_N, t;
    [|\psi(t)\ra, \qHcnc])dx'_n,\enskip n = 1, \ldots, N.
}{6b}
Eqs.~(6b) and (5a) \e{together} are the linear superposition construc%
tion of the probability vector current density $\vjc$ for any canoni%
cal Hamiltonian operator $\qHcnc$ whose classical Hamiltonian function
antecedent is $\Hclc$---see Eq.~(5a) for the way that $\Hclc$ is util%
ized.

We now turn to the best-known Hamiltonian operator, namely $|\qvp -
\vp_0(\qvx, t)|^2/(2m)$, for which \e{direct application of} Eq.~(3c)
suffices to extract the corresponding probability vector current den%
sity divergence $\del_\vx\dt\vj(\vx, t;[|\psi(t)\ra, |\qvp - \vp_0(
\qvx, t)|^2/(2m)])$~[1], which we  abbreviate as $\del_\vx\dt\vj$ for
convenience.  It turns out that this divergence can simply be \e{al%
gebraically manipulated} into the explicit form of a \e{divergence
operator} $\del_\vx\dt\>$ \e{acting on a certain vector field}~[1],
\e{without actually making use of} Eq.~(6b).  We can then inquire
whether the resulting ``textbook'' probability vector current densi%
ty $\vj(\vx, t; [|\psi(t)\ra, |\qvp - \vp_0(\qvx, t)|^2/(2m)])$~[1]
that is specifically thus obtained by \e{algebraic manipulation} of
its divergence $\del_\vx\dt\vj$ \e{in fact adheres to the three stipu%
lations used to derive} Eq.~(6b), namely homogeneous linearity in 
$\del_\vx\dt\vj$, no additional constants beyond the ones intrinsic to
$\del_\vx\dt\vj$, and the maximum possible symmetry.

\sct{Do the three postulated stipulations hold for the best-known spe%
cial example?}
\noindent
The key to applying Eq.~(3c) to the Hamiltonian operator $|\qvp -
\vp_0(\qvx, t)|^2/(2m)$ is obviously to work out its Eq.~(3c) \e{core
part}, $\la\vx||\qvp - \vp_0(\qvx, t)|^2/(2m)|\psi(t)\ra$,
\re{
    \la\vx||\qvp - \vp_0(\qvx, t)|^2/(2m)|\psi(t)\ra =
    \la\vx|(|\qvp|^2 - \qvp\dt\vp_0(\qvx, t) - \vp_0(\qvx, t)\dt\qvp +
    |\vp_0(\qvx, t)|^2)|\psi(t)\ra/(2m) =\cr
    \la\vx|(|\qvp|^2 + i\h\del_{\qvx}\dt\vp_0(\qvx, t) -
    2\vp_0(\qvx, t)\dt\qvp + |\vp_0(\qvx, t)|^2)|\psi(t)\ra/(2m) =\cr
    \bigl[-\h^2\del^2_\vx\psi(\vx, t) +
    2i\h\vp_0(\vx, t)\dt\del_\vx\psi(\vx, t) +
    (i\h\del_\vx\dt\vp_0(\vx, t) + |\vp_0(\vx, t)|^2)\psi(\vx, t)\bigr]/(2m),
}{7a}
where $\psi(\vx, t)\eqdf\la\vx|\psi(t)\ra$, and of course there is
also its complex conjugate $\bar\psi(\vx, t)\eqdf\la\psi(t)|\vx\ra$.
We now obtain the divergence $\del_\vx\dt\vj(\vx, t; [|\psi(t)\ra,
|\qvp - \vp_0(\qvx, t)|^2/(2m)])$, which we abbreviate as $\del_\vx\dt
\vj$ for convenience, by merely putting Eq.~(7a) into Eq.~(3c).  But
we \e{also} find that with \e{considerable additional algebraic mani%
pulation} that result can be explicitly presented as a \e{divergence
operator} $\del_\vx\dt\>$ acting on a certain vector field,
\re{
    \del_\vx\dt\vj = \Re\!\lf[(2i/\h)\la\psi(t)|\vx\ra\la\vx||\qvp -
    \vp_0(\qvx, t)|^2/(2m)|\psi(t)\ra\rt] =\cr
    m^{-1}\Re\!\lf[-i\h\bar\psi(\vx, t)(\del^2_\vx\psi(\vx, t))
                 -2\bar\psi(\vx, t)(\vp_0(\vx, t)\dt\del_\vx\psi(\vx, t))
                 -|\psi(\vx, t)|^2(\del_\vx\dt\vp_0(\vx, t))\rt] =\cr
    (2m)^{-1}[\bar\psi(\vx, t)((-i\h)\del^2_\vx\psi(\vx, t)) +
              ((i\h)\del^2_\vx\bar\psi(\vx, t))\psi(\vx, t)] -\cr
    m^{-1}[\bar\psi(\vx, t)(\vp_0(\vx, t)\dt\del_\vx\psi(\vx, t)) +
           (\vp_0(\vx, t)\dt\del_\vx\bar\psi(\vx, t))\psi(\vx, t) +
           |\psi(\vx, t)|^2(\del_\vx\dt\vp_0(\vx, t)] =\cr
    (2m)^{-1}\del_\vx\dt[\bar\psi(\vx, t)((-i\h)\del_\vx\psi(\vx, t)) +
              ((i\h)\del_\vx\bar\psi(\vx, t))\psi(\vx, t)] -\cr
    m^{-1}[(\vp_0(\vx, t)\dt\del_\vx|\psi(\vx, t)|^2) +
           |\psi(\vx, t)|^2(\del_\vx\dt\vp_0(\vx, t))] =\cr
    \del_\vx\dt[\bar\psi(\vx, t)((-i\h)\del_\vx\psi(\vx, t)) +
              ((i\h)\del_\vx\bar\psi(\vx, t))\psi(\vx, t)
              -2\vp_0(\vx, t)|\psi(\vx, t)|^2]/(2m).
}{7b}
Here the divergence \e{operator} $\del_\vx\dt\>$ \e{has been simply
factored out of the divergence expression} $\del_\vx\dt\vj$ which fol%
lows from Eq.~(3c).  Therefore it is abundantly clear that the \e{re%
sulting} ``textbook'' probability vector current density~[1],
\re{
\vj(\vx, t; [|\psi(t)\ra, |\qvp - \vp_0(\qvx, t)|^2/(2m)])\eqdf\cr
[\bar\psi(\vx, t)((-i\h)\del_\vx\psi(\vx, t)) +
((i\h)\del_\vx\bar\psi(\vx, t))\psi(\vx, t)
- 2\vp_0(\vx, t)|\psi(\vx, t)|^2]/(2m),
}{7c}
is \e{homogeneously linear} in its Eq.~(3c) divergence $\del_\vx\dt
\vj$ and also that it has \e{no additional constants} beyond those
that are \e{intrinsic} to $\del_\vx\dt\vj$.  Inspection of its
Eq.~(7c) expression also reveals this particular probability vector
current density to be highly symmetric; indeed its N-dimensional form
is the \e{completely symmetric generalization of its one-dimensional
case}.  Thus this ``textbook'' best-known special example of a proba%
bility vector current density~[1] \e{definitely adheres to the three
stipulations which underlie} Eq.~(6b).

One naturally wonders which subclass of the class of canonical Hamil%
tonian operators $\qHcnc$ has members which all emulate $|\qvp - \vp_0
(\qvx, t)|^2/(2m)$ insofar as having the property that the \e{diver%
gence operator} $\del_\vx\dt\>$ \e{can simply be explicitly algebrai%
cally factored out of} $\dvjc$.  An important clue for working out
that subclass is the fact that this property is manifest in the clas%
sical limit, where the single-particle vector current density can
readily be shown through \e{precisely} such explicit algebraic factor%
ization to be equal to the singular classical single-particle scalar
density times the classical particle velocity.  Interestingly, the
straightforward quantization of the singular classical single-particle
scalar density turns out to be equal to the quantum particle-position
projection operator in the Heisenberg picture, and, of course, the
\e{quantum} probability scalar density \e{is simply an expectation
value of that particle-position projection operator}.  Thus it isn't
greatly surprising that there exists \e{a subclass of the canonical
Hamiltonian operators} $\qHcnc$ for which, \e{in Ehrenfest-theorem
style}, the probability vector current density turns out to be equal
to the \e{expectation value} of the \e{quantization} of the just-men%
tioned result for the \e{classical} single-particle vector current
density, namely the singular classical single-particle scalar density
times the classical particle velocity, a result that, \e{exactly as in
the classical case}, is arrived at for this ``Ehrenfest subclass'' of
the canonical Hamiltonian operators \e{by the explicit algebraic fac%
torization} of the \e{divergence operator} $\del_\vx\dt\>$ out of
$\dvjc$.

\sct{The classical vector current density and an Ehrenfest-like theo%
rem}
\noindent
As is familiar from its electromagnetic application~[4], the \e{clas%
sical} single-particle scalar density is given by the singular expres%
sion,
\re{
    \rho_\cs(\vy, t) = \dl^{(N)}\!\!(\vx_\cs(t) - \vy),
}{8a}
which, \e{irrespective} of the value of t, satisfies,
\re{
    \int\rho_\cs(\vy,t)\,d^N\vy = 1,
}{8b}
in precise analogy to the probability-conservation property of the
\e{quantum} single-particle  probability scalar density $\la\psi(t)|
\vy\ra\la\vy|\psi(t)\ra$, which satisfies,
\re{
    \int\la\psi(t)|\vy\ra\la\vy|\psi(t)\ra\,d^N\vy =
    \la\psi(t)|\psi(t)\ra = 1.
}{8c}
Therefore it is plausible that there exists a classical single-par%
ticle vector current density $\vj_\cs(\vy,t)$ which satisfies the
equation of continuity in conjunction  with $\rho_\cs(\vy,t)$,
\re{
    d\rho_\cs(\vy,t)/dt + \del_\vy\dt\vj_\cs(\vy, t) = 0.
}{8d}
For Eq.~(8d) to hold it must be the case that,
\re{
    \del_\vy\dt\vj_\cs(\vy, t) = -d\rho_\cs(\vy,t)/dt =
    -\{\dl^{(N)}\!\!(\vx_\cs(t) - \vy),\, \Hclcs\} =\cr
    -\bigl(\del_{\vx_\cs(t)}\dl^{(N)}\!\!(\vx_\cs(t) - \vy)\bigr)\dt
    \del_{\vp_\cs(t)}\Hclcs =\cr
    \bigl(\del_\vy\dl^{(N)}\!\!(\vx_\cs(t) - \vy)\bigr)\dt
    \del_{\vp_\cs(t)}\Hclcs =\cr
    \del_\vy\dt\bigl[\dl^{(N)}\!\!(\vx_\cs(t) - \vy)\>
    \del_{\vp_\cs(t)}\Hclcs\bigr] =
    \del_\vy\dt\bigl[\dl^{(N)}\!\!(\vx_\cs(t) - \vy)\>
    d\vx_\cs(t)/dt\bigr], 
}{8e}
where $\{\enskip, \enskip\}$ denotes the classical Poisson bracket,
and we have used Hamilton's first classical equation of motion, $d\vx_
\cs(t)/dt = \del_{\vp_\cs(t)}\Hclcs$.  Eq.~(8e) clearly shows that the
\e{divergence operator} $\del_\vy\dt\>$ \e{explicitly algebraically
factors out} of the divergence of the classical single-particle vector
current density $\del_\vy\dt\vj_\cs(\vy, t)$, thus yielding for the
classical single-particle vector current density $\vj_\cs(\vy, t)$
\e{itself}, 
\re{
   \vj_\cs(\vy, t) = \dl^{(N)}\!\!(\vx_\cs(t) - \vy)\>\del_{\vp_\cs(t)}\Hclcs
   = \rho_\cs(\vy, t)\>d\vx_\cs(t)/dt.
}{8f}
From Eq.~(8f) we see that the classical single-particle vector cur%
rent density is equal to \e{the classical single-particle scalar den%
sity times the classical particle velocity}, which is a simple, physi%
cally graphic result that turns out to have significant resonance in
canonical quantum mechanics as well.

The quantum/classical linkage arises \e{both} from the Eq.~(2) rela%
tion between commutator and Poisson brackets \e{and} from the fact
that the \e{quantization} of the classical single-particle scalar den%
sity $\rho_\cs(\vy, t) = \dl^{(N)}\!\!(\vx_\cs(t) - \vy)$ is the Heis%
enberg-picture version of the quantum position projection operator $|
\vy\ra\la\vy|$, \e{whose expectation value in the state} $|\psi(t)\ra$
\e{is the quantum probability scalar density} $\la\psi(t)|\vy\ra\la\vy
|\psi(t)\ra$.  This aspect of the quantum position projection operator
$|\vy\ra\la\vy|$ can be obtained from the fact that its application to
any position eigenstate vector $|\vx\ra$ yields,
\re{
    |\vy\ra\la\vy|\vx\ra = \dl^{(N)}\!\!(\vy - \vx)|\vy\ra =
    \dl^{(N)}\!\!(\vx - \vy)|\vx\ra,    
}{9a}
which is \e{identical} to what the application of $\dl^{(N)}\!\!(\qvx
- \vy)$ to that position eigenstate vector $|\vx\ra$ yields, namely,
\re{
    \dl^{(N)}\!\!(\qvx - \vy)|\vx\ra = \dl^{(N)}\!\!(\vx - \vy)|\vx\ra,
}{9b}
and the fact that, in the \e{canonical} quantum regime, the set $\{|
\vx\ra\}$ of position eigenstate vectors is \e{complete}.  Therefore,
the quantum probability scalar density $\la\psi(t)|\vy\ra\la\vy|\psi
(t)\ra$ has the \e{alternate} expression,
\re{
    \la\psi(t)|\vy\ra\la\vy|\psi(t)\ra = \la\psi(t)|\dl^{(N)}\!\!(\qvx - \vy)
    |\psi(t)\ra,
}{9c}
which is easily shown to produce the following \e{alternate form of}
Eq.~(3c),
\re{
    \dvjcy = (i/\h)\la\psi(t)|[\dl^{(N)}\!\!(\qvx - \vy),\, \qHcnc]
             |\psi(t)\ra.
}{9d}
Eq.~(9d), in turn, implies the following alternate form of the combi%
nation of Eqs.~(6b) and (5a),
\re{
    \bigl(\vjcy\bigr)_n =\cr (1/N){\dy\int_{-\infty}^{y_n}}(i/\h)\la\psi(t)|
    [\dl^{(N)}\!\!(\qvx - (y_1,\ldots,y'_n,\ldots, y_N)),\, \qHcnc]
    |\psi(t)\ra dy'_n,\enskip n = 1, \ldots, N.
}{9e}
Eq.~(9e) does not, however, provide information on the consequences of
undertaking Fourier decomposition of the canonical Hamiltonian opera%
tor $\qHcnc$, as Eq.~(5a) does.

Much less general than Eq.~(9e), but doubtless of greater physical in%
terest, is the probability vector current density result for the sub%
class of canonical Hamiltonian operators $\qHcnc$ which satisfy a par%
ticular ``Ehrenfest'' requirement, namely that,
\re{
    [\dl^{(N)}\!\!(\qvx - \vy),\, \qHcnc] = i\h\,\Q{\,
    \{\dl^{(N)}\!\!(\vx - \vy),\, \Hclc\}\,}\, ,
}{10a}
where, of course,
\re{
    \qHcnc = \,\Q{\,\Hclc\,}\, .
}{10b}
From Eq.~(2) it is immediately apparent that canonical Hamiltonian op%
erators of the form,
\re{
    \qHcnc = \,\Q{\,K([\vp], t) + V([\vx], t)\,}\, =
     K([\qvp], t) + V([\qvx], t),
}{10c}
do indeed \e{satisfy} the ``Ehrenfest'' requirement of Eq.~(10a).

Given a canonical Hamiltonian operator $\qHcnc$ which satisfies this
``Ehrenfest'' requirement, the substitution of Eq.~(10a) into Eq.~(9d)
permits a series of steps that are \e{analogous to those of the clas%
sical} Eq.~(8e), and that \e{likewise} result in the \e{explicit alge%
braic factorization} of the \e{divergence operator} $\del_\vy\dt\;$
out of the Eq.~(9d) expression for the divergence $\dvjcy$ of the
probability vector current density, thereby yielding the probability
vector current density $\vjcy$ \e{itself} as the \e{expectation value
of the quantization of the classical single-particle scalar density}
$\dl^{(N)}\!\!(\vx - \vy)$ \e{times the classical particle velocity}
$\del_\vp\Hclc$.  Carrying out this substitution of Eq.~(10a) into
Eq.~(9d), and then proceeding in analogy with Eq.~(8e), we obtain,
\re{
    \dvjcy = -\la\psi(t)|\,\Q{\,\{\dl^{(N)}\!\!(\vx - \vy),\,
    \Hclc\}\,}\,|\psi(t)\ra =\cr
    -\la\psi(t)|\,\Q{\,\bigl(\del_\vx\dl^{(N)}\!\!(\vx - \vy)\bigr)\dt
    \del_\vp\Hclc\,}\,|\psi(t)\ra =
    \la\psi(t)|\,\Q{\,\bigl(\del_\vy\dl^{(N)}\!\!(\vx - \vy)\bigr)\dt
    \del_\vp\Hclc\,}\,|\psi(t)\ra =\cr
    \la\psi(t)|\,\Q{\,\del_\vy\dt\bigl[\dl^{(N)}\!\!(\vx - \vy)\>
    \del_\vp\Hclc\bigr]\,}\,|\psi(t)\ra =\cr
    \del_\vy\dt\bigl[\la\psi(t)|\,\Q{\,\dl^{(N)}\!\!(\vx - \vy)\>
    \del_\vp\Hclc\,}\,|\psi(t)\ra\bigr],
}{10d}
where, because quantization and taking the divergence with respect to
the non-quantized vector variable $\vy$ are independent linear proces%
ses, we can extract the explicitly factored divergence operator $\del_
\vy\dt\>$ out of the quantization---and also, of course, out of the
expectation value with the state $|\psi(t)\ra$.  Therefore, for the
subclass of canonical Hamiltonian operators $\qHcnc$ which satisfy the
``Ehrenfest'' requirement of Eq.~(10a), the probability vector current
density is given by,
\re{
    \vjcy = \la\psi(t)|\,\Q{\,\dl^{(N)}\!\!(\vx - \vy)\>
    \del_\vp\Hclc\,}\,|\psi(t)\ra,
}{10e}
which is \e{the expectation value of the quantization of the classical
single-particle scalar density} $\dl^{(N)}\!\!(\vx - \vy)$ \e{times
the classical particle velocity} $\del_\vp\Hclc$.

Now in \e{addition} to those \e{particular} ``Ehrenfest-subclass''
canonical Hamiltonian operators which have the form given by Eq.~%
(10c), canonical Hamiltonian operators which have the form,
\re{
    \qHcnc = \,\Q{\,\vp\dt\vv_0(\vx, t)\,}\, =
    \hf(\qvp\dt\vv_0(\qvx, t) + \vv_0(\qvx, t)\dt\qvp),
}{10f}
\e{as well} satisfy the ``Ehrenfest'' requirement of Eq.~(10a), which
we now show in an abbreviated manner,
\de{[\dl^{(N)}\!\!(\qvx - \vy),\, \hf(\qvp\dt\vv_0(\qvx, t) + \vv_0(\qvx, t)
     \dt\qvp)] = i\h\bigl(\del_{\qvx}\dl^{(N)}\!\!(\qvx - \vy)\bigr)\dt
     \vv_0(\qvx, t) =\cr i\h\bigl(-\del_{\vy}\dl^{(N)}\!\!(\qvx - \vy)
     \bigr)\dt\vv_0(\qvx, t) = -i\h\del_\vy\dt\bigl[\dl^{(N)}\!\!(\qvx -
     \vy)\>\vv_0(\qvx, t)\bigr],}
and,
\de{\,\Q{\,\{\dl^{(N)}\!\!(\vx - \vy),\, \vp\dt\vv_0(\vx, t)\}\,}\, =
\,\Q{\,\bigl(\del_\vx\dl^{(N)}\!\!(\vx - \vy)\bigr)\dt\vv_0(\vx, t)\,}\, =
\,\Q{\,\bigl(-\del_\vy\dl^{(N)}\!\!(\vx - \vy)\bigr)\dt\vv_0(\vx, t)\,}\, =\cr
-\del_\vy\dt\,\Q{\,\bigl[\dl^{(N)}\!\!(\vx - \vy)\>\vv_0(\vx, t)\bigr]\,}\, =
-\del_\vy\dt\bigl[\dl^{(N)}\!\!(\qvx - \vy)\>\vv_0(\qvx, t)\bigr].}
These results make it apparent that for this case as well, the diver%
gence operator $\del_\vy\dt\>$ explicitly algebraically factors out,
paving the way for the resulting probability vector current density to
be equal to the expectation value of the quantization of the classical
single-particle scalar density $\dl^{(N)}\!\!(\vx - \vy)$ times the
classical particle velocity $\vv_0(\vx, t) = \del_\vp \Hclc$.

Combining the special cases of Eqs.~(10c) and (10f), we see that can%
onical Hamiltonian operators $\qHcnc$ which adhere to the ``Ehren%
fest'' requirement of Eq.~(10a) have the form,
\re{
    \qHcnc = \,\Q{\,K([\vp], t) + \vp\dt\vv_0(\vx, t) + V([\vx], t)\,}\, =\cr
     K([\qvp], t) + \hf(\qvp\dt\vv_0(\qvx, t) + \vv_0(\qvx, t)\dt\qvp) +
     V([\qvx], t).
}{10g}
For this ``Ehrenfest subclass'' of the canonical Hamiltonian opera%
tors, the divergence operator $\del_\vy\dt\>$ \e{always explicitly al%
gebraically factors out} of the Eq.~(9d) expression $\dvjcy$ for the
divergence of the probability vector current density, as we see from
Eq.~(10d), and the consequent probability vector current density
$\vjcy$ \e{itself} is always given by \e{the expectation value of the
quantization of the classical single-particle scalar density} $\dl^
{(N)}\!\!(\vx - \vy)$ \e{times the classical particle velocity} $\del_
\vp\Hclc$, as we see from Eq.~(10e).

\vskip 1.75\baselineskip\noindent{\frtbf References}
\vskip 0.25\baselineskip

{\parindent = 15pt
\sk\item{[1]}
L. I. Schiff,
\e{Quantum Mechanics}
(McGraw-Hill, New York, 1955), pp.~23--24.
\sk\item{[2]}
S. K. Kauffmann,
Foundations of Physics {\bf 41},
805 (2011);
arXiv:0908.3755 [quant-ph],
(2009).
\sk\item{[3]}
S. K. Kauffmann,
Prespacetime Journal {\bf 1},
1249 (2010);
arXiv:0910.2490 [physics.gen-ph],
(2009).
\sk\item{[4]}
L. D. Landau and E. M. Lifshitz,
\e{The Classical Theory of Fields}
(Pergamon Press, Oxford, 1962).
}
\bye